\documentclass[english]{article}
\usepackage[T1]{fontenc}
\usepackage[latin9]{inputenc}
\usepackage{geometry}
\usepackage{refstyle}
\usepackage{amsmath}
\usepackage{amssymb}
\usepackage{graphicx}
\usepackage{esint}

\makeatletter

\AtBeginDocument{\providecommand\eqref[1]{\ref{eq:#1}}}
\AtBeginDocument{\providecommand\tabref[1]{\ref{tab:#1}}}
\AtBeginDocument{\providecommand\figref[1]{\ref{fig:#1}}}
\AtBeginDocument{\providecommand\subref[1]{\ref{sub:#1}}}
\providecommand{\tabularnewline}{\\}
\RS@ifundefined{subref}
  {\def\RSsubtxt{section~}\newref{sub}{name = \RSsubtxt}}
  {}
\RS@ifundefined{thmref}
  {\def\RSthmtxt{theorem~}\newref{thm}{name = \RSthmtxt}}
  {}
\RS@ifundefined{lemref}
  {\def\RSlemtxt{lemma~}\newref{lem}{name = \RSlemtxt}}
  {}

\makeatother

\usepackage{babel}
\begin{document}

\title{Effective Connectivity from Single Trial fMRI Data by Sampling Biologically
Plausible Models}

\author{H.C. Ruiz Euler, H.J. Kappen}
\maketitle
\begin{abstract}
The estimation of causal network architectures in the brain is fundamental
for understanding cognitive information processes. However, access
to the dynamic processes underlying cognition is limited to indirect
measurements of the hidden neuronal activity, for instance through
fMRI data. Thus, estimating the network structure of the underlying
process is challenging. In this article, we embed an adaptive importance
sampler called Adaptive Path Integral Smoother (APIS) into the Expectation-Maximization
algorithm to obtain point estimates of causal connectivity. We demonstrate
on synthetic data that this procedure finds not only the correct network
structure, but also the direction of effective connections from random
initializations of the connectivity matrix. In addition--motivated
by contradictory claims in the literature--we examine the effect of
the neuronal time scale on the sensitivity of the BOLD signal to changes
in the connectivity and on the maximum likelihood solutions of the
connectivity. We conclude with two warnings: First, the connectivity
estimates under the assumption of slow dynamics can be extremely biased
if the data was generated by fast neuronal processes. Second, the
faster the time scale, the less sensitive the BOLD signal is to changes
in the incoming connections to a node. Hence, connectivity estimation
using realistic neural dynamics time scale requires extremely high
quality data and seems infeasible in many practical data sets.
\end{abstract}

\section{Introduction}

In recent years, the field of neuroimaging has seen a rapidly increasing
interest in effective connectivity estimations from functional magnetic
resonance imaging (fMRI) data. Although this data acquisition method
is very powerful to investigate human brain function by identifying
brain regions that are active during perceptual or cognitive tasks,
fMRI time series are an indirect, delayed and blurred measurement
of the actual signal of interest, the latent neuronal activation of
a specific region of interest (ROI). Thus, estimating the underlying
connectivity is challenging. 

Roughly speaking there are two approaches to this problem based on
the distinction between functional and effective connectivity. The
first approach seeks to estimate the temporal correlations between
separable ROIs \cite{friston1994functional}. Examples of these, which
are known to correctly estimate the network structure, are partial
correlation, regularized inverse covariance and some Bayes net methods
\cite{smith2011network}. 

These symmetric measures, however, have no information about the directionality
of the connection, i.e. the influence that one node exerts over another.
This \textquotedbl{}effective\textquotedbl{} connectivity is in general
harder to estimate but it is often of great interest. We can distinguish
between two families of approaches; purely data-driven methods that
attempt to infer directionality directly from the time series using
statistical measures and methods based on dynamic models that seek
to find a forward model to fit the data. 

Purely data-driven methods fall generally into three classes. First,
lag-based methods, e.g. Granger causality \cite{granger1969investigating,goebel2003investigating,friston2013analysing},
are variations of the well-known auto-regressive models. In this framework
one considers the similarity between a pair of time series, one of
which is shifted in time. If the lagged time series helps predict
the zero-lagged time series, then a causal relation is inferred. A
second family of methods is based on the concept of conditional independence
and follow the ideas from structural learning in Bayes networks \cite{judea2000causality,ramsey2010six}.
The third class considers higher order statistics to infer causality.
For instance, Patel's $\tau$ approach measures the asymmetry between
the conditional probabilities of functionally connected nodes. If
the activation probability of a node $1$ given node $2$ is larger
than the activation probability of node $2$ given node $1$, this
is interpreted as a directed connection $2\rightarrow1$ \cite{patel2006bayesian}.
Although many of these methods are widely used, the estimation of
directed connectivity proves difficult with these methods \cite{smith2011network}.

Dynamic model methods frame the inference problem as a state-space
model (SSM). The latent states are the dynamic degrees of freedom
of the process underlying the data. Hence, in general, estimating
the parameters of the model requires also the estimation of the hidden
process, which is in most cases analytically infeasible and some approximations
are needed. Once the latent process estimation is addressed, one typically
uses the Expectation-Maximization (EM) algorithm in either its Maximum-Likelihood
(ML) or variational Bayes versions \cite{sarkka2013bayesian,ryali2011multivariate}. 

For fMRI data, model-based methods usually distinguish themselves
in the approximations they make of the dynamic system or the latent
process estimation (E-step). For instance, in \cite{ryali2011multivariate}
the fMRI time series is modeled as a linear convolution of the bi-linear
latent system with the canonical hemodynamic response function (HRF)
and its time derivative \cite{penny2005bilinear}. This approximation
allows using the Kalman smoother for the E-step. 

On the contrary, dynamic causal modeling (DCM) uses a biologically
plausible nonlinear model for the hidden process. Here, the neuronal
activity is given as a bi-linear system coupled to the Balloon model
describing the nonlinear relation between blood flow and volume \cite{buxton1998dynamics,mandeville1999evidence,friston2000nonlinear}.
In addition, the fMRI observations are modeled as a nonlinear function
of the hidden variables. Hence, to deal with the nonlinearities of
the biological model, DCM resorts to either a Volterra expansion of
the dynamic system or the generalized filtering method \cite{friston2003dynamic,friston2010generalised,friston2011network}. 

In addition, DCM uses a Gaussian approximation of the posterior over
the parameters and its mean and covariance are estimated using EM
update rules. For this, the adjacency matrix is specified beforehand,
effectively imposing a strong prior on the connectivity of the system.
Using the log-evidence, the proposed models are scored to find the
best model amongst them \cite{friston2011network}. Although DCM has
been widely used, concerns about this approach have spurred a discussion
about its feasibility and validity due to the combinatorial explosion
of the network structure and the apparent inability of the scoring
procedure to distinguish between generally accepted networks and randomly
generated ones \cite{lohmann2012critical,breakspear2013dynamic,friston2013model}. 

In this article, we present an alternative method for connectivity
estimation. Our approach is similar to DCM in the sense that it uses
the same model but differs in two important ways. First of all, in
the E-step the full posterior over the latent process is estimated
using an optimal control approach that was first introduced in \cite{ruiz2017particle}.
Second, in the M-step the connectivity is optimized without prior
assumptions as in \cite{friston2011network,friston2003dynamic}. The
Monte Carlo estimates involved in these steps are the only approximations
required and it is proven to work for nonlinear deconvolution of fMRI
time series \cite{ruizeuler2017nonlindeconv}.

We show using synthetic data that the connectivity estimates obtained
are close to the ground truth and that adding a small L1-regularization
on the connections is beneficial to obtain sparse estimates that generalize
better on unseen time series. Furthermore, the proposed method obtains
estimates from single event fMRI time series that are robust against
random initializations of the connectivity matrix.

In addition, we study the sensitivity of the BOLD signal depending
on the neuronal time scale and the effect of this parameter on the
connectivity estimates. This analysis is motivated by the different
claims about the value of the neuronal time scale in the literature.
For instance, \cite{smith2011network} used a time constant resulting
in a mean neural lag of 50 ms and it is argued that this value is
towards the upper limit of observed lags in general. Contrary to this
claim is the assumption made in the DCM literature, where it is argued
that the scale ought to be of order 1 s \cite{friston2003dynamic,friston2011network}.

We observe a significant effect of the neuronal time scale on the
connectivity estimates and a remarkable lower sensitivity of the BOLD
signal to changes in the connectivity for fast neuronal dynamics.
Hence if the underlying neuronal processes are fast, the quality of
the data must be significantly higher to obtain reliable, unbiased
estimates of the effective connectivity.

\section{Method }

\subsection{Modeling fMRI Data\label{sub:Modeling-fMRI-Data}}

Similar to DCMs \cite{friston2003dynamic}, we consider the fMRI forward
model as a network of $m$ regions $z=(z_{1},z_{2},\dots,z_{m})$
that follow stochastic dynamics given by

\begin{equation}
dz_{t}=A(Cz_{t}+u(z_{t},t)+BI_{t})dt+\sqrt{A}\sigma_{z}dW_{t}\label{eq:neuronal-activity-dyn}
\end{equation}

where $dW_{t}\sim N(0,dt)$ is a $m$-dimensional Wiener process with
variance%
\footnote{In all our simulations we used this discretization step unless stated
otherwise.%
} $dt=0.01$ and $\sigma_{z}>0$. The parameter $A$ sets the time
scale of the neuronal response, while the connectivity matrix $C$
has diagonal with $-1$ and off-diagonal elements $\left|C_{ij}\right|\leq1,i\neq j$.
This form assumes that the fastest scale is the within-node temporal
decay. The term $\sqrt{A}$ in the diffusion ensures that the stationary
distribution remains invariant under changes in the time scale $A$.
Contrary to DCMs, the input strength has been rescaled such that the
stationary point of the system with non-zero constant input is independent
of $A$. Hence, the inverse time scale $A$ determines only how fast
the neuronal system follows the input.

Notice that we define the process in \eqref{neuronal-activity-dyn}
to have an unknown function $u(z,t)$. This function is the importance
sampling controller learned with APIS. In general, $u(z,t)$ can be
any parametrized function \cite{kappen2015adaptive}, but in this
paper it is chosen to have the simple form $u(z,t)=az+b_{t}$ with
$a$ a constant and $b_{t}$ a time varying function. We assume that
the external input $I_{t}$ and its strength $B$ are known.

Each node's activity $z_{i}$ is coupled to a nonlinear deterministic
system modeling the hemodynamic transformation. For each node $i$
there are two Hemodynamic equations \cite{friston2000nonlinear} 
\begin{align}
ds_{i} & =\left(\epsilon z_{i}-\frac{s_{i}}{\tau_{s}}-\frac{f_{i}-1}{\tau_{f}}\right)dt\label{eq:hemodyn-eqs}\\
df_{i} & =s_{i}dt\nonumber 
\end{align}
 and two equations of the Balloon model%
\footnote{For clarity in the notation, we denote the exponent of the volume
fraction $1/\alpha$ simply as $\alpha$.%
} \cite{buxton1998dynamics},

\begin{align}
dq_{i} & =\frac{1}{\tau_{0}}\left(f_{i}\frac{1-(1-E_{0})^{1/f_{i}}}{E_{0}}-v_{i}^{\alpha-1}q_{i}\right)dt\label{eq:balloon-model}\\
dv_{i} & =\frac{1}{\tau_{0}}\left(f_{i}-v_{i}^{\alpha}\right)dt.\nonumber 
\end{align}

The BOLD signal change is given by

\begin{equation}
y_{i}(t)=B(q_{i},v_{i}|\theta)+\sigma_{y}dW_{y}\label{eq:BOLD-signal}
\end{equation}
where $B(q_{t},v_{t}|\theta):=V_{0}\left[k_{1}\left(1-q_{t}\right)+k_{2}\left(1-\frac{q}{v}\right)+k_{3}\left(1-v\right)\right]$,
$dW_{y}\sim N(0,1)$ and $\theta$ denotes all parameters of the system.
For simplicity, we assume the same hemodynamic transformation for
all nodes since our focus is on connectivity estimates. The parameters
are found in \cite{friston2000nonlinear}. A summary of all the values
used in this article is given in \tabref{Parameters-of-the-BOLD-trafo},
unless a different value is stated explicitly. For now, we restrict
our attention to slow neuronal processes with a neuronal lag of $A=1\ Hz$,
which is around the typical values in the literature on effective
connectivity, e.g. \cite{friston2003dynamic,friston2011network}.
The input strength $B$ is chosen such that the resulting amplitude
of BOLD responses are around 2-4\%.

\begin{table}
\centering{}%
\begin{minipage}[t]{1\columnwidth}%
\begin{center}
\begin{tabular}{|c|c|c|c|c|c|c|c|}
\hline 
Parameter & Value & Parameter & Value & Parameter & Value & Parameter & Value\tabularnewline
\hline 
\hline 
$A$ & 1 Hz & $\sigma_{z}$ & $10^{-3}$ & $B$ & $2.5$ & $\mu_{z,0}$ & $0$\tabularnewline
\hline 
\hline 
$\epsilon$ & 0.8 & $E_{0}$ & 0.4 & $\alpha^{-1}$ & 0.32 & $\sigma_{z,0}$ & $\sigma_{z}/\sqrt{2}$\tabularnewline
\hline 
$\tau_{s}$ & 1.54 & $k_{1}$ & $7E_{0}$ & $V_{0}$ & 0.018 & $\sigma_{s,0}$ & TBD\tabularnewline
\hline 
$\tau_{f}$ & $2.44$ & $k_{2}$ & $2$ & $\mu_{s,0}$ & $0$ & $\sigma_{f,q,v,0}$ & TBD\tabularnewline
\hline 
$\tau_{0}$ & $1.02$ & $k_{3}$ & $2E_{0}-0.2$ & $\mu_{f,q,v,0}$ & $1$ & $\sigma_{y}$ & $0.002$\tabularnewline
\hline 
\end{tabular}
\par\end{center}

\caption{Parameters for the neural dynamics (top row) and for the BOLD transformation
(bottom rows). \label{tab:Parameters-of-the-BOLD-trafo}}
\end{minipage}
\end{table}

In addition, we consider for the prior over the initial states in
each node $x_{i,0}=(z_{i,0},s_{i,0},f_{i,0},q_{i,0},v_{i,0})$ a normal
distribution%
\footnote{Notice that due to the small discretization step $dt$ and noise levels
used here, the log-transformation of the hemodynamic variables was
not required \cite{stephan2008nonlinear}. Nevertheless, it is straightforward
to use this transformation in our procedure.%
} with mean $\mu_{0}=(\mu_{z,0},\mu_{s,0},\mu_{f,0},\mu_{q,0},\mu_{v,0})$
and a covariance given by a diagonal matrix with entries $\sigma_{(z,s,f,q,v),0}^{2}$
set to be the variance of the stationary distribution induced by the
Ornstein-Uhlenbeck process in \eqref{neuronal-activity-dyn} when
$I_{t}=u(z,t)=0$. Hence, the variance for $p(z_{0})$ is set to $\sigma_{z,0}=\sigma_{z}/\sqrt{2}$.
Since all hemodynamic variables are deterministic, their variance
is estimated by forward sampling.

Now that the model has been described, we proceed with a more detailed
explanation of the parameter inference procedure given by the EM-algorithm.

\subsection{EM-algorithm for Time Series}

The Expectation-Maximization algorithm \cite{dempster1977maximum}
is an iterative method to find maximum-likelihood or maximum a posteriori
(MAP) estimates. There are several applications to state space models
(SSM), e.g. \cite{schon2011system}. In its general form, its objective
is to maximize a lower bound of the marginal likelihood $p(y_{0:T}|\theta)=\int dz_{[0:T]}p(z_{[0:T]},y_{0:T}|\theta)$
where $y_{0:T}=\{y(t_{k})|k=0,\dots,K\}$ is the time series, $\theta$
the set of parameters to estimate and $z_{[0:T]}=\{z_{t}|t\in[0,T]\}$
are the hidden (continuous) processes underlying the observations.
For any probability density $q(z_{[0:T]})$ we get from Jensen's inequality
\[
\log\left[p(y_{0:T}|\theta)\right]\geq-D_{KL}\left[q(z_{[0:T]})||p(z_{[0:T]},y_{0:T}|\theta)\right]
\]
where $D_{KL}\left[q(z_{[0:T]})||p(z_{[0:T]},y_{0:T}|\theta)\right]$
is the Kullback-Leibler (KL) divergence. Hence, the variational density
$q(z_{[0,T]})$ must be optimized to tighten this inequality. It turns
out that the optimal variational density $q(z_{[0:T]})$ for a fixed
value of the parameters $\theta=\tilde{\theta}$ is precisely the
posterior $p(z_{[0:T]}|y_{0:T},\tilde{\theta})$ \cite{neal1998view}.
Hence, given an initial value $\tilde{\theta}$, the objective is
\begin{equation}
\theta^{*}=\underset{\theta}{\text{argmax}}\left[-D_{KL}\left[p(z_{[0:T]}|y_{0:T},\tilde{\theta})||p(z_{[0:T]},y_{0:T}|\theta)\right]\right]\label{eq:Objective-Func}
\end{equation}

This algorithm entails two steps in each iteration. First, the E-step
is used to obtain expectations with respect to the posterior over
the latent process $z_{[0,T]}$. Then, the M-step is a gradient update
to maximize the objective function with respect to the parameters
in question.

\subsubsection{E-step: APIS}

Given the above model, we can sample from the posterior using APIS.
This adaptive importance sampling method samples $N$ trajectories,
or particles, by forward integration of (\ref{eq:neuronal-activity-dyn})-(\ref{eq:balloon-model})
initialized with $u(z,t)=0$. Then, a total cost $S_{\xi}$ is assigned
to each hidden process $\xi$ \cite{girsanov1960transforming,kappen2005linear,kappen2012optimal}
via 
\[
S_{\xi}=-\int_{0}^{T}dt\log\left[g(y_{0:T}|z_{[0,T]}^{(\xi)})\right]-\frac{1}{2}\frac{A}{\sigma_{z}^{2}}(u_{t}^{(\xi)})^{2}-\frac{\sqrt{A}}{\sigma_{z}}u_{t}^{(\xi)}dW_{t}^{(\xi)}
\]
where $dW_{t}^{(\xi)}$ denotes the noise realization of process $\xi$,
$u_{t}^{(\xi)}=u(z_{t}^{(\xi)},t)$ and $g(y_{0:T}|z_{[0,T]}^{(\xi)})$
is the likelihood with respect to the latent process $z_{[0,T]}^{(\xi)}$
given by \eqref{BOLD-signal}. We call the first term of $S_{\xi}$
the \textquotedbl{}state cost\textquotedbl{} of particle $\xi$. The
negative of the total cost exponentiated gives the unnormalized importance
weight used in the Monte Carlo estimates to learn the control function
$u(z,t)$. After $u(z,t)$ is updated, it is used to sample again
forward particles. These steps are repeated until the effective sample
size (ESS) converges or reaches a predefined value. For details on
APIS we refer the reader to \cite{ruiz2017particle}.

APIS not only increases the efficiency of sampling, but it also minimizes
the mean cost $S$ of the particle ensemble. This ensures that the
posterior neuronal and BOLD signals give the best possible fit, even
if there is a strong mismatch between the model and the ground truth
\cite{ruizeuler2017nonlindeconv}. In addition, the control appears
in the computation of the gradients of the KL divergence, giving reliable
Monte Carlo estimates in the M-step.

For a 3 node network, we use $N=20000$ particles over 500 iterations
to ensure that the scheme has sufficient time to bootstrap the E-step
and converge to the MAP connections in the M-step. Depending on the
initialization, the estimations can take more or less iterations,
but we find that this number allows for convergence in most cases.
A learning rate of $\eta_{E}=0.1$ achieves a fast bootstrapping while
maintaining the estimations of the control signals stable. For the
annealing threshold we use $\gamma_{E}=0.02-0.03$, corresponding
to 2-3\% of effective samples used for the estimations during bootstrapping.

\subsubsection{M-Step: Gradient Ascent in the Connectivity}

In the M-step, we maximize \eqref{Objective-Func} with respect to
the connectivity matrix $C$ and assume that all other parameters
have correct values. Since $C$ is assumed to be time independent,
this optimization requires estimates of $z_{t}$ at all time. Thus,
we wish to maximize at the $n$-th iteration, 
\[
Q_{n}(C)=\sum_{t=dt}^{T}\mathbb{E}\left\{ \log\left[p(z_{t}|z_{t-dt},C)\right]|y_{0:T},C^{(n)}\right\} 
\]
where the expectation is over the posterior latent process. Using
\eqref{neuronal-activity-dyn} and the conditional independence of
the components $z_{t}^{i}$ given $z_{t-dt}$, the above results in
\[
Q_{n}(C)=\sum_{t=dt}^{T}\sum_{i=1}^{m}\mathbb{E}\left\{ \frac{-1}{2\sigma_{z}^{2}dt}\left(dz_{t}^{i}-F_{i}(z_{t},C)dt\right)^{2}|y_{0:T},C^{(n)}\right\} 
\]
where $F(z_{t},C)=A(Cz_{t}+BI_{t})$. With $\partial_{ik}F:=\partial F/\partial C_{ik}=Az_{t}^{k}$
and $dz_{t}-F(z_{t},C)dt=Au_{t}dt+\sqrt{A}\sigma_{z}dW_{t}$ we finally
obtain 
\[
\partial_{ik}Q_{n}=\sum_{t=dt}^{T}\mathbb{E}\left\{ \frac{A}{\sigma_{z}^{2}}\left(Au_{t}^{i}dt+\sqrt{A}\sigma_{z}dW_{t}^{i}\right)z_{t}^{k}|y_{0:T},C^{(n)}\right\} .
\]

This gradient is used to update the connectivity at each iteration,
\[
C_{ik}^{(n+1)}=C_{ik}^{(n)}+\eta_{M}\partial_{ik}Q_{n}
\]
where $\eta_{M}$ is the learning rate. 

Note that both the E- and the M-step involve optimization. In the
E-step we optimize the importance sampler for given $C^{(n)}$. In
the M-step, we optimize $C^{(n)}$. In practice, we find that it is
beneficial to wait until the ESS in the E-step is sufficiently high.
This is ensured by imposing a threshold $\gamma_{M}$ on the ESS above
which the connectivity is updated. Empirically, it is found that $\gamma_{M}\in[2\gamma_{\text{E}},5\gamma_{E}]$
and a learning rate of $\eta_{M}=0.01$ work well. The latter is usually
found such that after each update, the ESS takes only few iterations
to reach $\gamma_{M}$ again. Naturally, there is a trade-off between
large updates of the connectivity and the stability of the ESS. In
addition, we use momentum with a rate of $\kappa=0.9$ to improve
the gradient ascent procedure. 

Remember that the within-node time scale is assumed to be the fastest
time scale in the network and all nodes have the same value. Hence,
we fix the diagonal elements of $C$ to  $-1$. Although they are
not optimized, the above procedure can be readily extended to estimate
these elements as well.

The initial random matrices for $n=0$ are generated with the following
procedure similar to \cite{lohmann2012critical}. First, on-edges
are sampled with probability $p_{1}=0.5$. This gives a random graph
that defines the adjacency matrix, i.e. the directed connections among
the ROIs. Then, the strength of these connections are randomly chosen
from a uniform distribution on a small interval $[c,d]$ where $\left|c\right|,d\leq1$.
Here, the bound on the strength are chosen to be $c=-0.5$ and $d=0.5$.
This ensures that the sampled matrix has negative eigenvalues with
very high probability, which is important to guarantee the stability
of the latent process. The resulting matrix is accepted if all eigenvalues
are negative, otherwise, the procedure is repeated.

\subsection{Studying the Effects of the Neuronal Time Scale on Effective Connectivity
Estimates}

The use of slow dynamics for the neuronal processes underlying fMRI
data is argued in \cite{friston2011network}, but it stands in contrast
to the case studied in \cite{smith2011network} with a neural lag
of approximately 50 ms, corresponding to $A=20\ Hz$. This is a significant
difference in the assumptions made on the generative process of the
data, hence, it is important to understand the consequences of these
assumptions for effective connectivity estimates. For this, we study
two important effects that depend on the time scale of the neuronal
activity, the sensitivity of the BOLD response to changes in the connectivity
and the maximum-likelihood solutions under different assumptions of
the time scale.

As a measure of sensitivity, the difference in the mean and variance
of the posterior BOLD between two significant different models is
computed. We call this difference the \textquotedbl{}sensitivity\textquotedbl{}
of the mean $s_{\mu}$ and variance $s_{\sigma}$ respectively. Hence,
if the change in the response is large when moving from one connectivity
model to another, we say that the system is highly sensible to changes
in the connectivity. On the contrary, small changes in the BOLD response
will make the discrimination between connectivity models more difficult. 

In addition, the inverse time scale $A$ not only affects the sensitivity
of the BOLD response, but also its overall shape and delay. This could
have significant effects on the connectivity estimates. Thus, we study
the solutions depending on the assumed neuronal time scale given that
the underlying generative process has fast neuronal activation, i.e.
we consider a model mismatch between the ground truth inverse time
scale $A_{GT}$ and the assumed one for the reconstruction of the
network $A$. 

For simplicity, we consider the deterministic system because we find
that the variance of the posterior contributes marginally to the neg.
log-likelihood ($s_{\sigma}\ll s_{\mu}$) and the posterior mean follows
the same dynamics as the deterministic system for sufficiently small
noise levels. Hence, since the only randomness is in the observations,
we do not use the EM approach described above. Instead, we use a brute
force search on the neg. log-likelihood landscape. This approach requires
the discretization of the connectivity space and, for each grid point,
the integration of the full system on the time interval $[0,T]$.

In this analysis, consider for the ground truth a chain network $I\rightarrow1\rightarrow2\rightarrow3$
with connectivity matrix $C_{GT,c}$ and input $I$. To lower the
computational effort in finding the connections we restrict the problem
by fixing all connections but two, say $\left(C_{31},C_{32}\right)$.
The brute force approach even in this restricted scenario is computationally
too expensive to study the grid search solutions for a large number
of time series. Hence, the search on this plane is constrained further
to the line $C_{31}=-gC_{32}+h$ on which, given the input, the fix
point of the neuronal system is invariant to changes in the connectivity.
This constrain gives $g=(C_{GT,c})_{23}$ and $h=g\cdot(C_{GT,c})_{32}$. 

To see this, consider the fix point solution $z^{*}$ of \eqref{neuronal-activity-dyn}
with $u(z,t)=0,\ \sigma_{z}=0$. For node 3, the dependence of $z_{3}^{*}$
on the connections $\left(C_{31},C_{32}\right)$ is given by $z_{3}^{*}=[gC_{32}+C_{31}]z_{1}^{*}$,
where $z_{1}^{*}$ is the fix point of the activity in node 1. Keeping
the input fixed sets the value of $z_{1}^{*}$, so $z_{3}^{*}$ will
be completely determined by the proportionality factor $h=gC_{32}+C_{31}$.
By setting $h=g\cdot(C_{GT,c})_{32}$, we ensure that the fix point
of the system remains constant at the value of the ground truth while
we vary the connections on this line, which we call the $z^{*}$-line.
Hence, the only relevant information to differentiate between connections
is in the transients given by the values on this line. We will denote
estimates obtained by the brute force search on the $z^{*}$-line
as grid search connections to distinguish them from the EM approach
described above. For a validation on this procedure we refer the reader
to the appendix.

We use the grid search on two studies on the effect of the neuronal
time scale on the grid search connections and the measurement precision
needed to obtain reliable connectivity estimates. In these studies,
we perform a brute force search on the $z^{*}$-line as follows. Given
a value $C_{32}$, we integrate the system to obtain the BOLD response,
which is used to compute the log-likelihood and find the grid search
solution of $1000$ noise realizations, first keeping $\sigma_{y}$
fixed and varying $A$, then keeping $A$ fixed to the correct value
$A_{GT}$ and varying $\sigma_{y}$. 

Notice that the variance observed in this analysis is clearly an under
estimate of the true variance of the grid search solutions in the
full connectivity space, because we restricted ourselves to a single
line $C_{31}=-gC_{32}+h$ in this space. However, the results of this
analysis capture already the consequences of fast neuronal dynamics
for connectivity estimates.

\section{Results}

\subsection{Single Trial Estimates of Effective Connectivity from Slow Neuronal
Activation}

In the following analysis we consider two connectivity structures
with different topologies with no bidirectional connection, a chain
network $C_{GT,c}$ and a triangle network $C_{GT,\Delta}$. In both
cases, we consider an external input to node 1 given by a box-car
function $I_{t,1}=\Theta[t-t_{on}]\Theta[t_{off}-t]$ where $(t_{on},t_{off})=(3.2,4.70)$
seconds and $\Theta$ is the Heaviside function. To generate the data
from both examples $C_{GT,c}$ and $C_{GT,\Delta}$, the systems are
integrated forward $T=16$ seconds with the initial state set to the
mean value $\mu_{0}$. In each case, the response signals of the nodes
are subsampled with a $TR=0.4$ seconds, resulting in 3 time series
each with 41 observations corrupted by Gaussian noise ($\sigma_{y}=0.002$).
This noise level is chosen to be 5-8\% of the BOLD change, roughly
the same as in \cite{friston2011network}.

We illustrate how the EM-algorithm together with APIS obtains the
correct directed edges. Since sparse networks are assumed, we use
an L1-regularization term $\Sigma_{i,j}\lambda_{L1}\left|C_{ij}\right|$
in \eqref{Objective-Func}.

\begin{figure}
\begin{centering}
\includegraphics[scale=0.7]{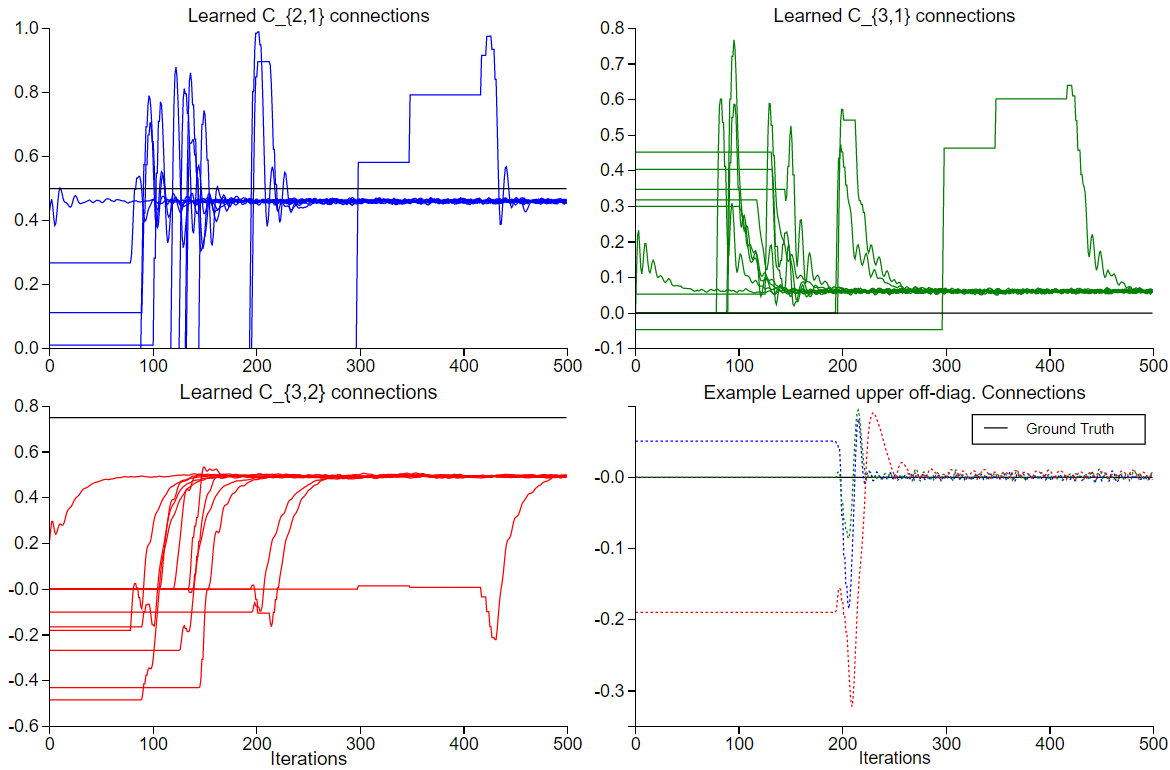}
\par\end{centering}

\caption{MAP-solutions for chain network: The figure shows the consistency
of the estimation procedure against random initializations of the
connectivity matrix. Each line represents the estimation procedure
over iterations of a randomly initialized example. The regularization
strength is set to $\lambda_{L1}=0.05$. All runs converged to the
same matrix $C_{\lambda_{L1}=0.05}^{c}$. The estimated connections
are not identical to the ground truth because of the finite data used
in reconstruction. However, the qualitative structure of the network,
including which connections are present and their directionality,
is correctly reconstructed. On the lower right panel only an example
of the upper off-diagonal connections is shown. In all cases they
are estimated to vanish with very high precision. \label{fig:3ROIs_Chain_A1Hz_Random-initialized-matrices} }
\end{figure}

For the first example, we consider the chain network with connections
\[
C_{GT,c}=\left(\begin{array}{ccc}
-1 & 0 & 0\\
0.5 & -1 & 0\\
0 & 0.75 & -1
\end{array}\right).
\]
In \figref{3ROIs_Chain_A1Hz_Random-initialized-matrices}, the connectivity
estimation for the chain network $C_{\lambda_{L1}=0.05}^{c}$ is shown
over iterations. In this case, we choose a regularization of $\lambda_{L1}=0.05$.
Each graph in the panels corresponds to a different random initialization
of the connectivity matrix. At the beginning, the connections do not
change because the sampler must learn an appropriate controller to
bootstrap the procedure. Once the controller is estimated with sufficient
accuracy, the ESS surpasses a threshold and the connectivity is updated
via gradient ascent, hence the sudden jumps in the connections. After
a few jumps, the learning procedure stabilizes and the connections
are updated at each iteration. In most cases, it takes around 200-250
iterations to converge to the MAP solution. 

The ESS after convergence of most examples increased from less than
1\% to higher than 90\% of total particles used. The mean state cost
of the sampled process $\mathbb{E}\left(\log\left[g(y_{0:T}|z_{[0,T]})\right]\right)$
decreased in all cases from around 1000 to 68, meaning that the model
found fits the data well.

Notice that the randomly initialized matrices converged all to the
same values close to the ground truth. The connection $C_{31}^{c}$
is sufficiently weak and could be regarded as non existent since the
other estimated connections dominate by an order of magnitude. The
underestimation of the connection $C_{32}^{c}$ can be understood
from the L1-penalty in the M-step to obtain sparse solutions. This
regularization shrinks the connections by penalizing their absolute
value. In turn, this shrinkage is compensated by an increase in the
other incoming connection to node 3 to reduce the state cost, i.e.
$C_{31}^{c}$ becomes non-zero. This hints to a strong regularization.
To see the effect of strong regularization, we study the dependency
of the solutions on the regularization strength $\lambda_{L1}$.

\begin{figure}
\begin{centering}
\includegraphics[scale=0.8]{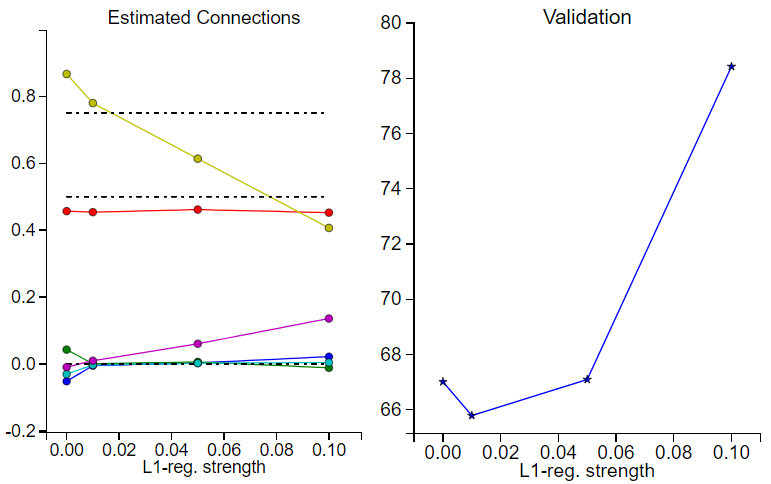}
\par\end{centering}

\caption{MAP-solutions for chain network vs L1-regularization strength: a small
amount of regularization $\lambda_{L1}=0.01$ improves the estimation
of the connectivity compared to the unregularized case. The mean validation
error of the BOLD signal over 20 new time series decreases compared
to the unregularized case but increases rapidly with stronger regularization
(right). Taking the connectivity at the minimum of the validation
error gives us a connectivity with a small error relative to the ground
truth connection (black dashed line, left panel) \label{fig:L1-regularization-strength-3ROIs_Chain}}
\end{figure}

For a fixed initialization of uncoupled nodes $C_{0}^{c}=-\mathbf{1}_{3\times3}$,
the connections of the chain network are estimated using different
values $\lambda_{L1}\in[0,0.1]$. The analysis shows--left panel of
\figref{L1-regularization-strength-3ROIs_Chain}--that the ground
truth connections (black dashed line) are consistently revealed across
a wide range of $\lambda_{L1}$-values. As expected, zero regularization
gives small non-zero contributions in edges that would not be allowed,
but still the ground truth connections are easy to differentiate.
Hence, although the L1-regularization is not crucial in this example,
we see that applying a small regularization helps shrinking the value
of non existent connections towards zero, which improves the estimate
of the network making it generalize better, as seen on the right panel
of \figref{L1-regularization-strength-3ROIs_Chain}, where the mean
validation error is shown. 

To estimate the validation error we generate 1000 new time series
using the ground truth connection and computed the mean negative log-likelihood
or sate cost of the response obtained from each connectivity. This
is a measure of how well our estimates fit the new data. We observe
a decrease for weak regularization $\lambda_{L1}=0.01$ vis-�-vis
the unregularized case and a strong increase in the validation error
for increasing $\lambda_{L1}$. From the minimum of this profile,
we conclude that a weak regularization is beneficial, in this case
using $\lambda_{L1}=0.01$ we obtain a solution pinpointing the ground
truth to
\[
C_{\lambda_{L1}=0.01}^{c}=\left(\begin{array}{ccc}
-1 & 0 & 0\\
0.45 & -1 & 0\\
0 & 0.78 & -1
\end{array}\right).
\]

\begin{figure}
\begin{centering}
\includegraphics[scale=0.7]{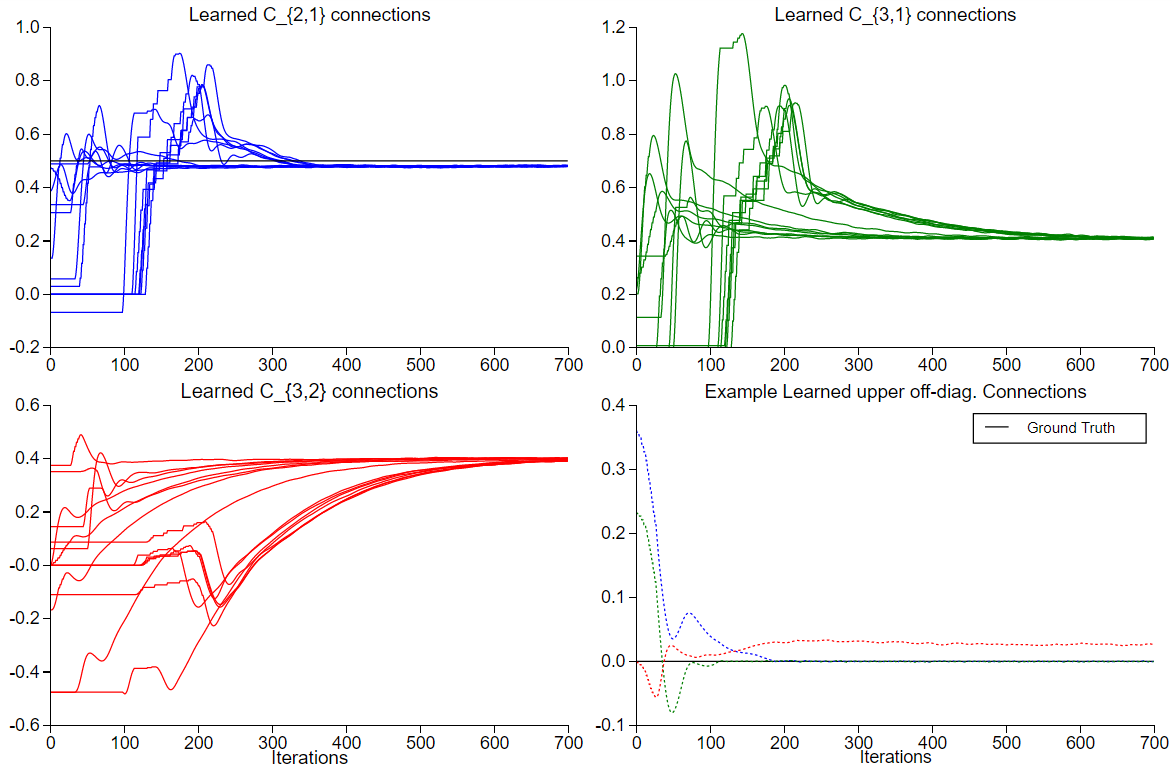}
\par\end{centering}

\caption{Triangle network with similar connection strengths as the chain network:
The MAP-solution is again robust against random initializations of
the connectivity and we can clearly distinguish between existent and
non-existent edges and directions.\label{fig:triangle-network-MAPestimations}}

\end{figure}

Finally, we proceed with the estimation of the triangle network 

\[
C_{GT,\Delta}=\left(\begin{array}{ccc}
-1 & 0 & 0\\
0.5 & -1 & 0\\
0.375 & 0.5 & -1
\end{array}\right)
\]
to show that we obtain again the ground truth structure with a small
regularization of $\lambda_{L1}=0.01$. In \figref{triangle-network-MAPestimations},
we observe a robust estimation against random initializations of the
connectivity matrix and a clear distinction between correct and incorrect
edges and directions. In this case, all estimated connections are
around
\[
C_{\lambda_{L1}=0.01}^{\Delta}=\left(\begin{array}{ccc}
-1 & 0 & 0\\
0.48 & -1 & 0.03\\
0.41 & 0.4 & -1
\end{array}\right).
\]

Interestingly, in this case the overestimated connection is $C_{23}^{\Delta}$
with the same order of magnitude as $C_{31}^{\Delta}$ in the previous
example. Again, we may vary $\lambda_{L1}$ to select its value by
cross validation. 

\subsection{Identifying Networks with Fast Neuronal Activations}

\subsubsection{BOLD Sensitivity to Changes in the Connectivity Diminishes with Fast
Neuronal Dynamics}

The connectivity estimation above assumed that the temporal scale
of the neuronal dynamics is of the order of a second. Here, we study
the sensitivity of the BOLD response to changes in the connectivity
as a function of $A$. We consider 3 different time scales $A=1,\ 5\text{ and }50$.
The neuronal noise level $\sigma_{z}=0.01$ is chosen such that the
prior process has a standard deviation of 0.7\%. The learning rate
of APIS has to be adapted for the different values of $A$, roughly
with the relation $\eta_{E}=0.1/A$. All other parameters of APIS
are kept fixed.

Given the data generated by the chain network $C_{GT,c}$, the posterior
statistics of the BOLD signal are estimated in two extreme cases using
APIS; one case defines the structure $3\leftarrow1\rightarrow2$ and
the other is close to the ground truth $C_{GT,c}$. The connectivity
matrices are chosen such that the only difference to the ground truth
are the connections into node 3, 
\begin{align*}
C_{1}=\left(\begin{array}{ccc}
-1 & 0 & 0\\
0.5 & -1 & 0\\
0.375 & 0 & -1
\end{array}\right); & \ C_{2}=\left(\begin{array}{ccc}
-1 & 0 & 0\\
0.5 & -1 & 0\\
0.06 & 0.63 & -1
\end{array}\right).
\end{align*}

\begin{figure}
\begin{centering}
\includegraphics{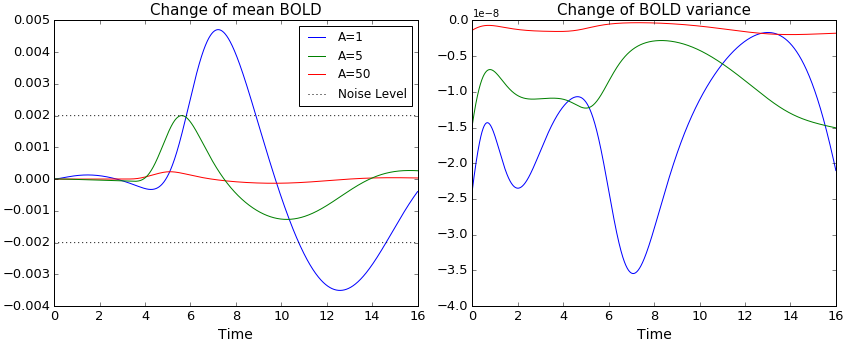}
\par\end{centering}

\caption{Sensitivity of the BOLD posterior to changes in the connectivity.
Notice that the change in the variance of the posterior BOLD (right)
is five orders of magnitude smaller than the change in the posterior
mean (left). Hence, faster neuronal activity implies less sensitivity
of the BOLD signal to changes in the connectivity. The dashed black
line is the noise level $\sigma_{y}=0.002$ of the data generated
with $C_{GT}^{c}$. \label{fig:Sensitivity-of-the-BOLD-to-connectivity}}

\end{figure}

In \figref{Sensitivity-of-the-BOLD-to-connectivity}, we show the
sensitivity of the BOLD response in node 3 when changing the connections
from $C_{1}$ to $C_{2}$. We appreciate two characteristics of the
posterior BOLD response. 

First, on the right panel, it is apparent that the variance of the
signal has little sensitivity to the connectivity, with a change several
orders of magnitude smaller than the change of the mean signal on
the left. This suggests that--for small neuronal noise--the contribution
of the variance to the identification of the connectivity is marginal
and we can focus on the mean signal, which is equal to the BOLD signal
of a deterministic system.

Second, on the left panel, for slow dynamics the amplitude of the
mean signal change is larger than the observation noise level $\sigma_{y}$.
However, already for $A=5$ the sensitivity decreases below the observation
noise and it becomes an order of magnitude smaller for large $A$.
Hence, the faster the neuronal dynamics, the less sensible the BOLD
response becomes to changes in the connectivity. The lack of sensitivity
for faster time scales implies flat likelihood functions that contain
no information to differentiate connectivity structures. Thus, $A$
should be sufficiently small such that the sensitivity is higher than
the observation noise. This seems to be a fundamental problem to reconstruct
connectivity.

One solution to this problem is more data, which effectively reduces
the measurement noise. Another, option is to reconstruct with small
$A$ and hope for the best. Thus, we study the grid search solutions
obtained for different inverse time scales $A$ by the brute force
search described before.

\subsubsection{Connectivity Estimates Depend on the Assumed Neuronal Time Scale}

\begin{figure}
\begin{centering}
\includegraphics[scale=0.5]{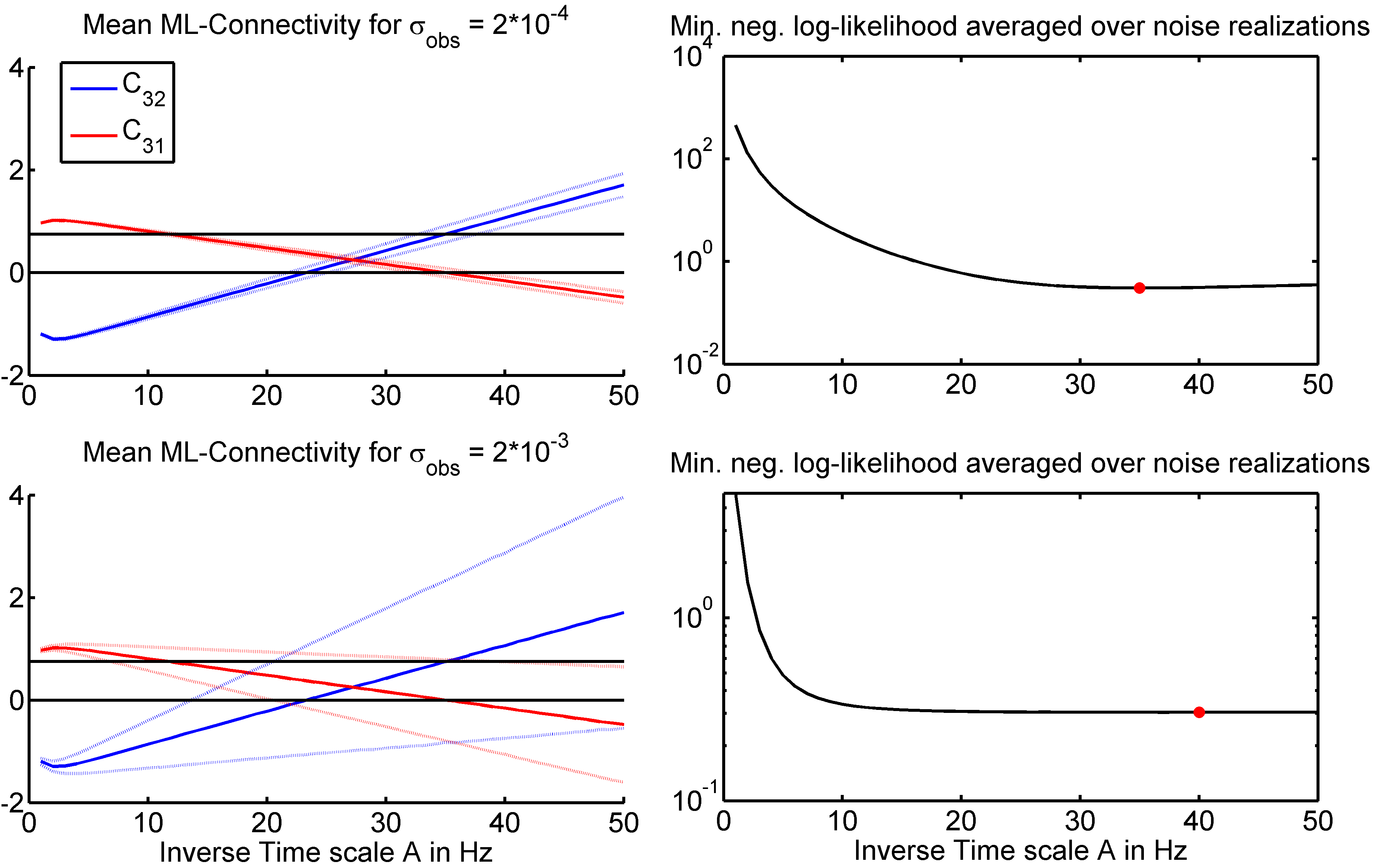}
\par\end{centering}

\caption{Connections obtained via grid search vs inverse time scale $A$ for
$\sigma_{y}=0.0002$ (top) and $\sigma_{y}=0.002$ (bottom). Left:
Mean (solid) and standard deviation (dashed) of grid search solutions
on $C_{31}=-gC_{32}+h$ over different realizations of the observation
noise. All connections are fixed besides the in-connections $C_{32},C_{31}$
to node 3. The horizontal black lines are the values of the ground
truth connections at $C_{32}=0.75$ and $C_{31}=0$. Right: The averaged
minimum of the neg. log-likelihood function for each inverse time
scale. The ground truth is given by the model described in \subref{Modeling-fMRI-Data}
but with $A_{GT}=35$. The red dot on the right panels represent the
minimum of the mean neg. log-likelihood profile. Due to the large
variance and the flat neg. log-likelihood the minimum in the bottom
case was estimated around $A=40$. \label{fig:ML-estimates-vs-time-scale-A}}
\end{figure}

We study now the connectivity estimates as a function of the inverse
neuronal time scale $A$. For each value of $A$, we find the connections
from 1000 different time series with the ground truth given by the
chain network $C_{GT,c}$ with the same parameters as before but $A_{GT}=35$.
Figure \ref{fig:ML-estimates-vs-time-scale-A} shows the mean and
standard deviation of these estimates. Taking the amplitude of $s_{\mu}$
in \figref{Sensitivity-of-the-BOLD-to-connectivity} as a rough upper
limit for the noise level that allows connectivity estimations, the
analysis is performed with two noise levels, $\sigma_{y}=2\times10^{-4}$
(on the top panels) and $\sigma_{y}=2\times10^{-3}$ (on the bottom
panels).

Upon examination of the top panels, we reinforce our expectation that
with sufficient precision, the variance of the grid search solutions
is small. Interestingly, the inverse time scale $A$ biases the connectivity
estimates in a significant way. Nevertheless, it is possible to obtain
in principle the correct connectivity if the time scale is jointly
estimated because there is a clear minimum in the negative log-likelihood
profile (red dot on the right panel). 

On the contrary, with a noise level an order of magnitude higher,
the identification of connections is problematic. Although the bias
of the estimations is the same as above, the variance of the grid
search solutions increases much faster with the inverse time scale
because of the flat profiles of the neg. log-likelihood function.
Thus, the grid search solutions spread across a wide range of values,
making it difficult to pinpoint the connectivity. As a consequence,
for $A\geq15\ Hz$ identifying the correct connections will not be
possible, even after learning the inverse time scale. This can be
appreciated by the flat profile of the mean neg. log-likelihood at
the grid search solutions. 

Our analysis focuses on grid search solutions on the restricted connectivity
space $C_{31}=-gC_{32}+h$ for simplicity. However, a similar analysis
on the bidirectional connections results also in an extreme bias in
the connectivity estimates due to the wrong time scale. This can be
appreciated by inspecting the neg. log-likelihood landscape of a bidirectional
connection with a mismatch in $A$ (not shown). As before, if the
data is generated by a fast process but one assumes a slow process,
the landscape is distorted and barely changes across different noise
realizations and values of $\sigma_{y}$. The distorted landscape
features a minimum around $(C_{32},C_{23})=(1.3,-1.3)$, although
the data was generated using $(C_{32},C_{23})=(0.75,0)$. 

\begin{figure}
\begin{centering}
\includegraphics[scale=0.4]{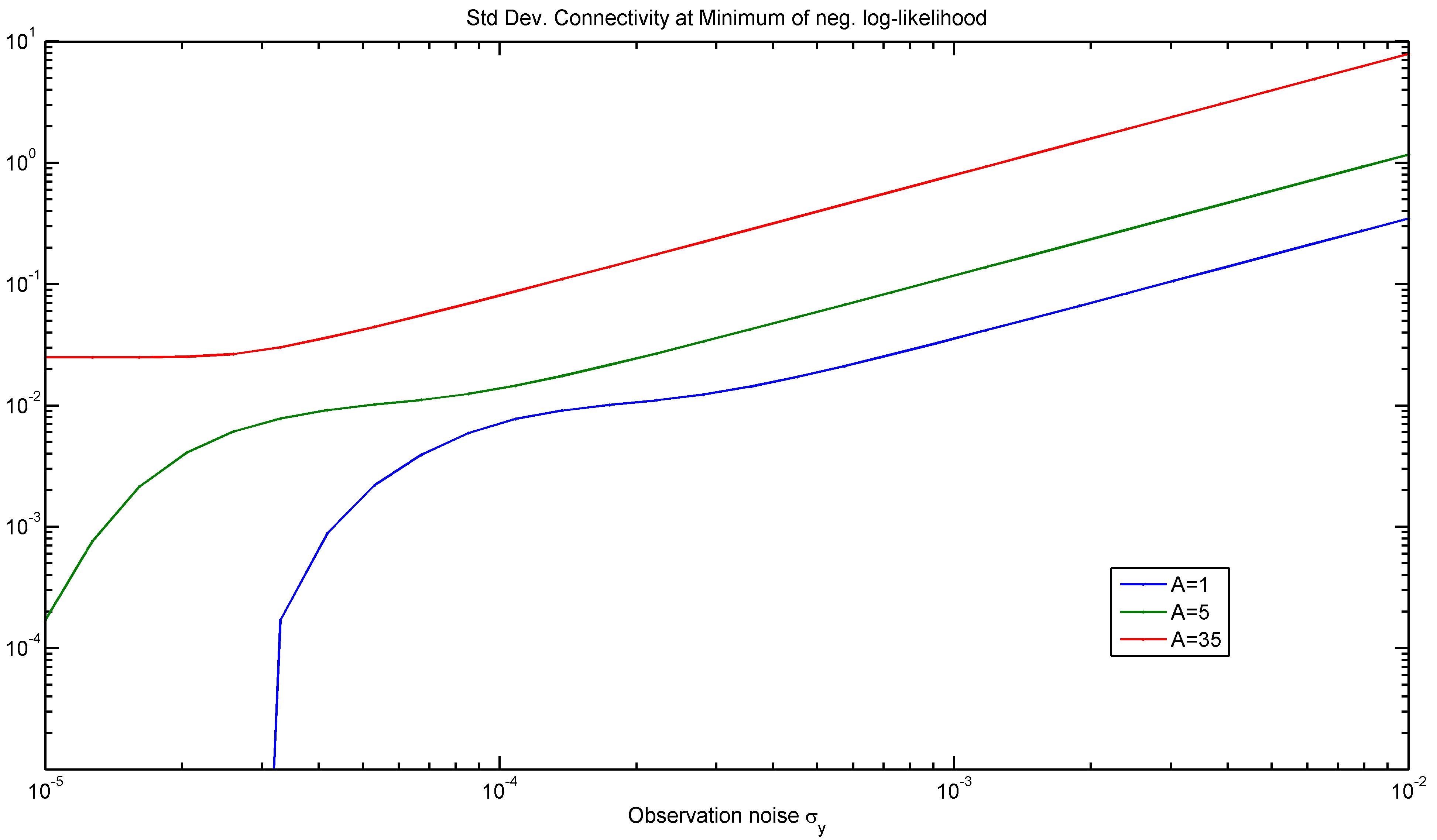}
\par\end{centering}

\caption{Standard deviation of grid search solutions for connection $C_{32}$
depending on the observation noise. The increase in precision needed
to obtain reliable connectivity estimates is a direct result of the
insensitivity of the BOLD signal to changes in the connectivity.
Notice the log-scaling of both axes. \label{fig:Standard-deviation-MLsolutions_vs_obsNoise}}

\end{figure}

Finally, we study the variance of the connectivity estimates as a
function of the observation noise $\sigma_{y}$. For this, data is
generated with $A_{GT}=1,\ 5\text{ and }35$ and the brute force search
is performed without model mismatch. As expected from \figref{ML-estimates-vs-time-scale-A},
the mean connectivity remains close to the ground truth values for
all noise levels (not shown). 

Figure \ref{fig:Standard-deviation-MLsolutions_vs_obsNoise} shows
the standard deviation of the grid search connections $C_{32}$. We
observe that the noise level allowed for connectivity estimation depends
on the inverse time scale of the neuronal dynamics. There are two
effects here. First and most intuitive, the variance of the grid search
connections increases with the observation noise. Notice the linear
increase for large values in this log-log-graph. Second but more important,
fast time scales in the neuronal signal exacerbate this increase at
least an order of magnitude due to the insensitivity of the BOLD signal
to changes in the connectivity. Notice how the dependency flattens
by lowering the noise, first for $A=1$, then for $A=5$ and finally
for $A=35$ in the extreme high precision regime $\sigma_{y}\propto10^{-5}$,
where the standard deviation of the grid search solutions for $A=1$
and $A=5$ collapse. This shows that--even in this idealized case--to
resolve connections when the neuronal time scale is around $30\ ms$
we require at most $\sigma_{y}<10^{-3}$.

\section{Discussion}

In this paper we present a method to obtain point estimates of effective
connectivity from fMRI time series. This method is based on the Expectation-Maximization
algorithm with the E-step being performed by APIS on the same biologically
plausible model used in DCM. We show that APIS-EM obtains robust estimations
independent of random initializations of the connectivity matrix.
In addition, we analyze the effect of the neuronal time scale on estimates
of the effective connectivity. This analysis shows that if the underlying
neuronal dynamics are fast, the connectivity estimates will be extremely
biased if slow processes are assumed in the reconstruction. Hence,
jointly estimating the neuronal time scale and the connectivity is
important. However, due to the high insensitivity of the BOLD signal
for faster neuronal dynamics, this becomes increasingly more demanding
and requires higher quality data.

We conclude that robust reconstruction of the connectivity between
ROIs including the directionality requires much higher precision in
the data when assuming a realistic (fast) time scale for neuronal
dynamics. In addition, if a slow time scale is assumed (as in \cite{friston2003dynamic,friston2011network})
the inferred connectivity can be completely wrong if the generative
neuronal process is fast.

The adaptive importance sampler APIS is an efficient tool to compute
posterior estimates over the hidden processes. Importantly, the sampling
step is easily parallelizable such that one can lower the computational
burden involved in integrating the system of non-linear differential
equations while having a large amount of samples to ensure reliable
estimates. 

Furthermore, APIS and its generalization based on PICE \cite{kappen2015adaptive},
are flexible frameworks that allow an easy adaptation of the M-step
to learn the hemodynamic parameters by including noise in the hemodynamic
degrees of freedom. Hence, instead of an $m$-dimensional controller,
APIS would learn a $5m$-dimensional controller. Although the computational
burden increases in this case, we find empirically that the fully
controlled version behaves more stable in terms of the effective sample
size. Nevertheless, for simplicity we focused here on the reconstruction
of the neural network and assumed the hemodynamic system is known.
Hence, to maintain a lower computational cost, we chose the simplified,
under-actuated setting.

Since the proposed scheme obtains robust estimates from randomly initialized
connectivity matrices, it can be considered as a gradient-based exploratory
procedure able to change the initial connectivity to find better fitting
models. This addresses some of the concerns raised in \cite{lohmann2012critical}.
Naturally, better initializations improve performance. A natural choice
is the symmetric connectivity obtained from methods known to infer
the right undirected network, for instance partial correlation or
regularized inverse covariance. 

Moreover, the addition of modulatory inputs or the consideration of
more complex neuronal systems, e.g. firing rate models with non-linear
activation functions, is straightforward. Furthermore, extensions
of the proposed method to obtain Gaussian approximations of the posterior
over all parameters similar to DCM can be worked out. This makes the
combination of APIS and EM a flexible and efficient alternative to
DCMs.

Finally, a word of caution. The bias in the connectivity estimates
assuming the wrong time scale has significant consequences for the
estimation of the hemodynamic parameters because the resulting BOLD
signal is shifted roughly a second with respect to the ground truth.
If both sets of parameters are jointly estimated, this shift will
bias the hemodynamic parameters as well. On the contrary, fast neuronal
dynamics make the delays being completely determined by the hemodynamics.
This might lower the \textquotedbl{}interference\textquotedbl{} between
delays caused by the neuronal time scale and the ones caused by the
hemodynamics, possibly increasing identifiability of the hemodynamic
system. Hence, it is important to clarify the temporal scale of the
underlying processes, specially because this assumption has a major
impact on the required quality of the data to obtain correct directed
networks.

\section*{Acknowledgements }

This work was supported by the European Commission through the FP7
Marie Curie Initial Training Network 289146, NETT: Neural Engineering
Transformative Technologies.

\section*{Appendix: Validation of Grid Search on $z^{*}$-Line}

We validate the grid search approach by studying the neg. log-likelihood
landscape on $\left(C_{31},C_{32}\right)$ for two extreme cases with
and without the correct inverse time scale. We show graphically that
when there is a mismatch in $A$, the bias caused by the restriction
on the $z^{*}$-line is small compared to the bias caused by the wrong
time scale.

\begin{figure}
\begin{raggedright}
\includegraphics[scale=0.17]{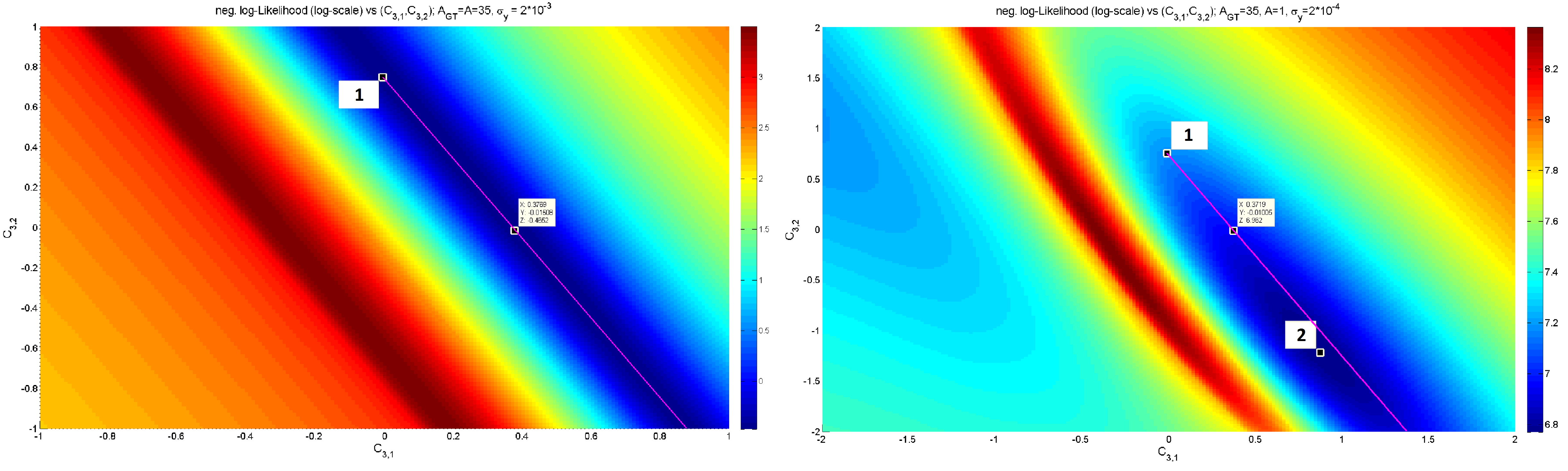}
\par\end{raggedright}

\caption{Validating the restriction of the search to the $z^{*}$-line (magenta
line). Negative log-likelihood in log-scale for two different cases
of the parameters $A$ and $\sigma_{y}$. The upper most data point
gives the position of the ground truth. Left: No mismatch in the time
scale of the model ($A=A_{GT}=35\ Hz$) and $\sigma_{y}=0.002$. Notice
the direction of the valley along the $z^{*}$-line. Right: There
is a mismatch between the ground truth time scale ($A_{GT}=35\ Hz$)
and the model ($A=1\ Hz$). Although the noise is small $\sigma_{y}=2\cdot10^{-4}$,
there is a wide valley in the form of an ellipse. Notice how the major
axis is parallel to the $z^{*}$-line. The minimum (lower data point)
falls in this case outside the line but the bias introduced by disregarding
the perpendicular direction is small compared to the overall bias
caused by the wrong time scale. \label{fig:Validating-the-restriction}}
\end{figure}

For the validation, data was generated using the chain network with
$A_{GT}=35$ and two different noise levels $\sigma_{y}=2\cdot10^{-3}$
and $\sigma_{y}=2\cdot10^{-4}$. Figure \ref{fig:Validating-the-restriction}
left panel shows the neg. log-likelihood on $\left(C_{31},C_{32}\right)$
for $\sigma_{y}=0.002$ estimated using the correct time scale. Although
the high noise level 'washes out' the $z^{*}$-line (magenta), there
is a clear direction along this line where the values are smallest.
Perpendicular to this direction the value increases. Thus, minima
lie in this case on the line. Due to the high noise level, the neg.
log-likelihood along this line is flat. For lower noise levels, the
valley becomes more and more narrow and the profile of the neg. log-likelihood
on this line shows a clear minimum.

On the right panel of \figref{Validating-the-restriction}, we consider
more precise data generated with $\sigma_{y}=2\cdot10^{-4}$ but compute
the neg. log-likelihood with a model mismatch in the time scale ($A=1$).
This distorts the landscape such that the low valued region around
the $z^{*}$-line transforms into an elliptic shaped valley with the
minimum (dot \#2) far away from the ground truth (dot \#1). Notice
how the major axis is parallel to the $z^{*}$-line and, although
the minimum does not lie exactly on the line, the error is small compare
to the bias caused by the wrong time scale. Thus, restricting the
search to the $z^{*}$-line is a good approximation for the purpose
of studying the grid search connections as a function of $A$.

\pagebreak{}

\bibliographystyle{plain}
\bibliography{C:/Users/HCRuiz/SURFdrive/PhD_backup/References/PhD_Bib_Database}

\end{document}